\documentclass[conference]{IEEEtran}

\usepackage[top=0.7in, left=0.68in, bottom=1.01in, right=0.68in]{geometry}

\IEEEoverridecommandlockouts
\usepackage{cite}
\usepackage{amsmath,amssymb,amsfonts}
\usepackage{algorithmic}
\usepackage{graphicx}
\usepackage{textcomp}
\usepackage[table]{xcolor}
\usepackage{cite}
\usepackage{balance}
\usepackage{bm}
\usepackage{algorithm}
\usepackage{algorithmic}
\usepackage{subcaption}

\DeclareMathOperator*{\argmax}{argmax}
\DeclareMathOperator*{\argmin}{argmin}

\begin{document}
	
	\bstctlcite{IEEEexample:BSTcontrol}
	
	\title{Full-Duplex Non-Coherent Communications for Massive MIMO Systems with Analog Beamforming
		\thanks{
			{This work was supported in part by Huawei Technologies Canada and in part by the Natural Sciences and Engineering Research Council of Canada.}
		}
	\vspace{-3.5ex}
	}
	
	\author{\IEEEauthorblockN{
			Asil Koc\IEEEauthorrefmark{1}, 
			Ahmed Masmoudi\IEEEauthorrefmark{3}, 
			Tho Le-Ngoc\IEEEauthorrefmark{1}}
		\IEEEauthorblockA{
			\IEEEauthorrefmark{1}Department of Electrical and Computer Engineering, McGill University, Montreal, QC, Canada \\
			{\IEEEauthorrefmark{3}IMT Atlantique, Lab-STICC, UMR CNRS 6285, F-29238 Brest, France}\\
			Email: asil.koc@mail.mcgill.ca,
			{ahmed.masmoudi@imt-atlantique.fr},
			tho.le-ngoc@mcgill.ca
		}
	\vspace{-6ex}
	}

	\maketitle
	
	\begin{abstract}
In this paper, a novel full-duplex non-coherent (FD-NC) transmission scheme is {developed} for massive multiple-input multiple-output (mMIMO) systems using analog beamforming (ABF).
	{We propose to use a} structured Grassmannian constellation for the non-coherent communications {that does not require channel estimation}.
Then, we design the transmit and receive ABF via the slow time-varying angle-of-departure (AoD) and angle-of-arrival (AoA) information, respectively.
	The ABF design targets maximizing the intended signal power while suppressing the strong self-interference (SI) occurred in the FD transmission.
Also, the proposed ABF technique only needs a single transmit and receive RF chain to support large antenna arrays, thus, it reduces hardware cost/complexity in the mMIMO systems.
	It is shown that the proposed FD-NC offers a great improvement in bit error rate (BER) in comparison to both half-duplex non-coherent (HD-NC) and HD coherent schemes.
	We also observe that the proposed FD-NC both reduces the error floor {resulted from the residual SI in FD transmission,} and provides {lower BER} compared to the FD coherent transmission.
	\end{abstract}
	
	\begin{IEEEkeywords}
		Full-duplex,
		non-coherent,
		massive MIMO,\hspace{2ex}
		analog beamforming,	
		self-interference cancellation,
		mmWave.
	\end{IEEEkeywords}

	\section{Introduction}
	{\IEEEPARstart{M}{illimeter} wave (mmWave) and massive multiple-input multiple-output (mMIMO) have been considered as enabling technologies for the fifth-generation (5G) and beyond wireless communication networks \cite{Uwaechia2020}.
		The shorter wavelengths in the mmWave frequency bands
		allow the large antenna array utilization in the practical mMIMO systems.
	By means of high beamforming gain, mMIMO with large antenna arrays can focus the signal energy in the intended direction to enhance the spectral efficiency and combat the limited scattering propagation in the mmWave channels.
		
		Although fully-digital beamforming (FDBF) is considered in the conventional MIMO systems \cite{Mass_MIMO_Precoding_Survey}, its application in mmWave mMIMO is not feasible due to the large channel estimation overhead and immense hardware cost/complexity.
	Since FDBF requires a dedicated expensive/power-hungry radio frequency (RF) chain for each antenna. 
		Recently, analog beamforming (ABF) is widely investigated for mmWave mMIMO systems, 
		which employs only a single RF chain to support large antenna arrays and performs the beamforming via low-cost phase-shifters at the RF-stage \cite{Mass_MIMO_Hyb_Survey,Analog_BF_mMIMO,FD_MIMO_ANALOG}. 

	Full-duplex (FD) communications can further extend the expected impacts of the mmWave mMIMO systems since it theoretically doubles the capacity via simultaneous transmission and reception over the same frequency band.
		Although FD is severely affected by the strong self-interference (SI), the recent developments in SI cancellation (SIC) techniques make it more practical 
		\cite{FD_Survey_2019, FD_Survey, FD_AntennaSep_74dB,ASIL_FD_SIC,FD_MIMO_ANALOG}.
	For instance, the experiment results in \cite{FD_AntennaSep_74dB} show that the antenna isolation based SIC can achieve up to $74$ dB cancellation on the strong near-field SI channel.
		Recently, a hybrid digital/analog beamforming (HBF) based SIC technique is developed for the FD mMIMO systems in \cite{ASIL_FD_SIC}, which only utilizes the slow time-varying angular information of the SI channel (i.e., the instantaneous SI channel knowledge is not required).
		Also, \cite{FD_MIMO_ANALOG} investigates the ABF design for FD mMIMO systems, where the authors assume the availability of full-size intended and SI channels to develop the transmit/receive ABF.		
		
	The majority of works on the beamforming design assume a priori channel state information (CSI) for coherent communications. However, the pilot-based CSI acquisition consumes a substantial portion of the channel coherence block especially for the large dimensional channels in mMIMO \cite{nonCohMasMIMO}.
		When a communication system does not rely on a priori CSI, so-called non-coherent, the entire coherence block can be exploited for data transmission.
	In \cite{cubeSplitConstellation}, half-duplex non-coherent (HD-NC) transmission using a structured Grassmannian constellation is studied for single-input multiple-output (SIMO) systems, where HD-NC enjoys the pilot-free transmission and provides better error performance compared to HD coherent scheme.

	
	
	
	
		In this paper, we develop a new full-duplex non-coherent (FD-NC) communications for mMIMO systems with ABF.
		A structured Grassmannian constellation is employed for the symbol detection without a priori CSI.
		The proposed transmit/receive ABF technique is based on the slow time-varying angular information, 
			where our targets are maximizing the intended signal power and suppressing the strong SI.
		It is shown that the proposed FD-NC achieves lower bit error rate (BER) than FD coherent, HD-NC and HD coherent schemes, especially for higher transmission rates.
	}
   	
   	The rest of this paper is organized as follows. 
	Section \ref{sec_SYS} introduces the system model and structured Grassmannian constellation.
	Section \ref{sec_Channel} describes the channel model.
	We propose the ABF design in Section \ref{sec_ABF}.
	After the illustrative	results in Section \ref{sec_RESULTS}, the paper is concluded in Section \ref{sec_CONC}.
	
	

	\section{\hspace{-1ex}System Model and Grassmannian Constellation}\label{sec_SYS}
	
	\subsection{System Model}\label{secsub_SYS}
\begin{figure}[!t]
		\centering
		\includegraphics[width = \columnwidth]{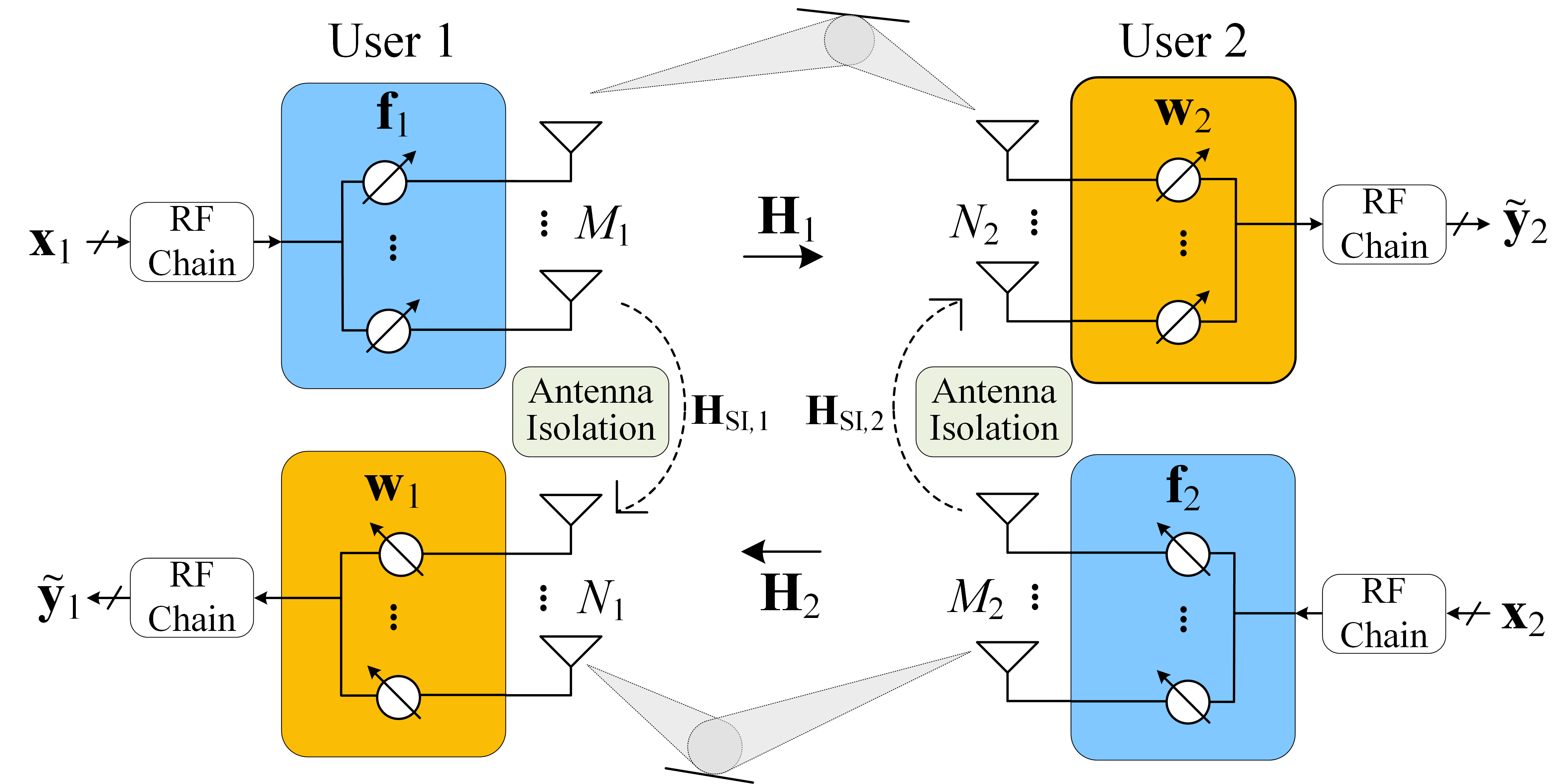}
		\caption{Full-duplex non-coherent mMIMO systems with ABF.}
		\vspace{-2ex}
		\label{fig1_System}
	\end{figure}

	Fig. \ref{fig1_System} represents the FD-NC mMIMO system model, where two users with large uniform linear arrays (ULAs) employ analog beamforming (ABF). 
	Specifically, the user $i\in{\left\{1,2\right\}}$ has a transmit ULA with $M_{i}$ antennas and a receive ULA with $N_{i}$ antennas. At the RF-stage, we utilize the low-cost phase-shifters to build the transmit ABF (i.e., ${\bf f}_i\in\mathbb{C}^{M_i}$) and receive ABF (i.e., ${\bf w}_i\in\mathbb{C}^{N_i}$). Hence, each user requires only a single transmit and receive RF chain to reduce the hardware cost/complexity.
		As seen from Fig. \ref{fig1_System}, an antenna isolation block is placed between transmit and receive ULAs to suppress the strong SI resulted from the FD transmission \cite{FD_AntennaSep_74dB}.
	
\begin{figure}[!t]
		\centering
		\includegraphics[width = \columnwidth]{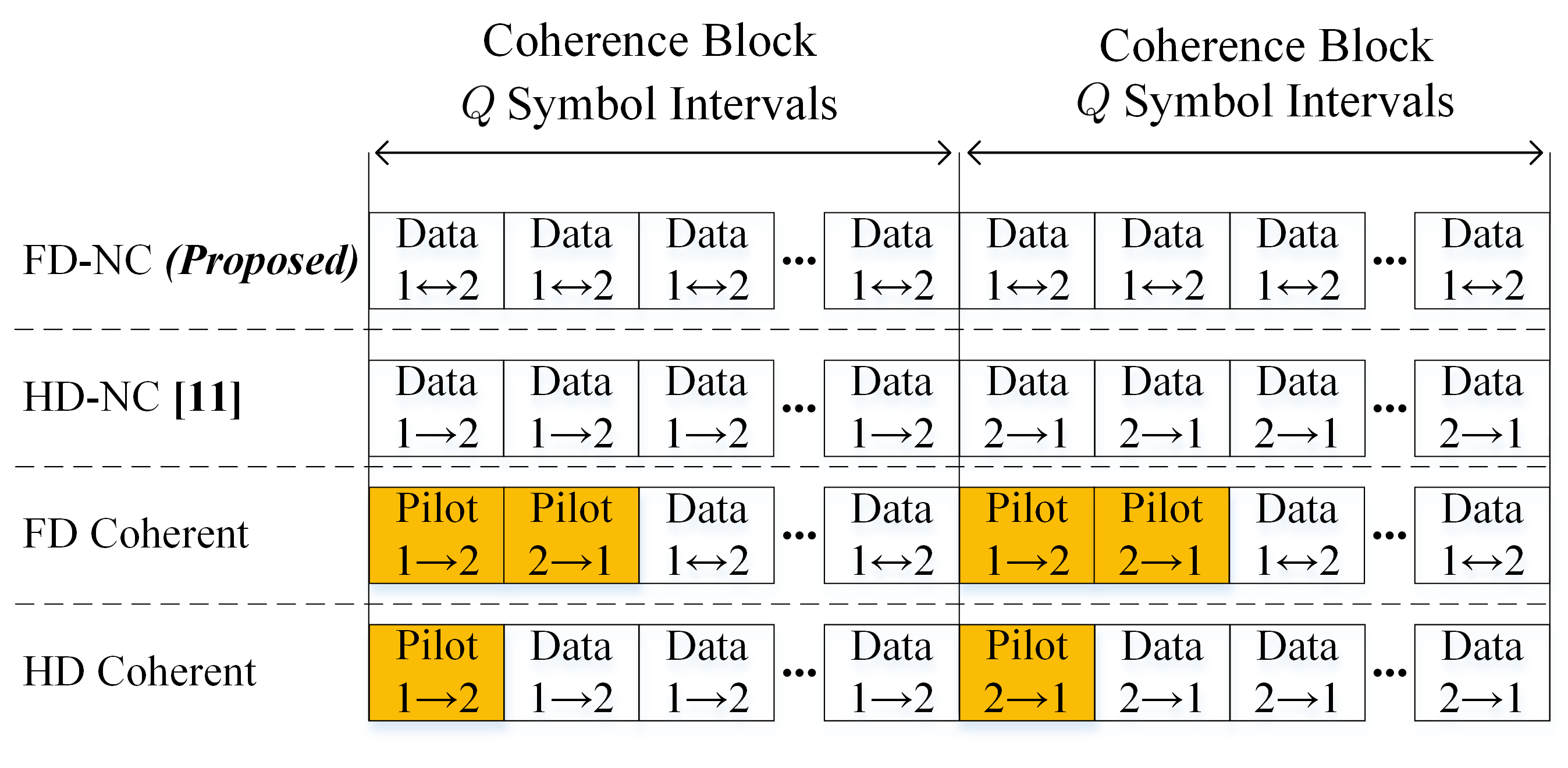}
		\caption{Transmission schemes.}
		\vspace{-2ex}
		\label{fig2_TransmissionSchemes}
	\end{figure}

The channel is assumed to be flat fading during coherence block of ${Q}$ symbol intervals. 
	Fig. \ref{fig2_TransmissionSchemes} demonstrates four possible transmission schemes based on the CSI requirement and {duplexing} mode of communication system.
First, the proposed new FD-NC transmission is applied without any pilot signaling, where each interval is utilized for two-way simultaneous data communications.
	Second, when HD non-coherent (HD-NC) transmission is utilized as in \cite{cubeSplitConstellation}, {the} entire coherence block is reserved for one-way data communications (i.e., either from user $1$ to user $2$, or from user $2$ to user $1$) without a priori CSI.
On the other hand, the coherent transmission schemes require to estimate the channel periodically in each coherence block. 
	Thus, for coherent transmission, we need to reserve at least one interval within each coherence block to send pilot symbols for channel estimation when using HD transmission, and two pilot intervals when using FD transmission.
Hence, $Q-1$ and ${Q}-2$ intervals are used for data communications in HD and FD coherent transmission, respectively, whereas all ${Q}$ intervals are exploited for data communications in both FD-NC and HD-NC.
	
	The transmitted signal during a coherence block by user $i$ is defined as
	${\bf S}_{i}={\bf f}_i {\bf x}_i^T\in\mathbb{C}^{M_i\times {Q}}$, where 
	${\bf x}_i\in\mathbb{C}^{{Q}}$ is the data symbol vector\footnote{
		In Section \ref{secsub_CS}, we present a structured Grassmannian constellation to construct the symbol vector ${\bf x}_i$ for non-coherent transmission.
	For the coherent communications, QAM can be utilized to generate each element of ${\bf x}_i$.}.
		Here, the transmitted signal is generated based on unit average power constraint per each interval. 
	Given a unit-norm transmit ABF vector (i.e., 
	${\bf f}_i^H{\bf f}_i=1$), the data symbol vector satisfies 
	$\mathbb{E}\big\{||{\bf S}_i||_F^2\big\}
	=
	\mathbb{E}\big\{||{\bf x}_i||^2\big\}
	=
	{Q}$.
	Before the receive ABF, the received signal matrix at user ${j\in\left\{1,2\right\}}$ during a coherence block is written as:
	\begin{equation}
		{\bf Y}_{j} 
		= 
		\sqrt{\rho}
		{\bf H}_i{\bf S}_{i} 
		+ 
		\sqrt{\rho}
		{\bf H}_{\textrm{SI},j}{\bf S}_{j} 
		+
		{\bf Z}_{j} \in\mathbb{C}^{N_j\times {Q}}, ~\forall j\ne i,
	\end{equation}
	where 
	$\rho\ge 0$
	is the transmit power,
	${\bf H}_i\in\mathbb{C}^{N_j\times M_i}$ 
	is the channel from user $i$ to user $j$,
	${\bf H}_{\textrm{SI},j}\in\mathbb{C}^{N_j\times M_j}$ 
	is the SI channel at user $j$,
	${\bf Z}_j \in\mathbb{C}^{N_j\times {Q}}$
	is the complex circularly symmetric Gaussian noise matrix {whose entries follow} the distribution of $\mathcal{CN}\left(0, \sigma_z^2\right)$.
		After applying the receive ABF at user $j$, the combined signal  is obtained as:
	\begin{equation}\label{eq_r_combined}
		\begin{aligned}
			\tilde{\bf y}_{j}^T
			&= 
			{\bf w}_j^T{\bf Y}_j
			\\
			&=
			\sqrt{\rho}
			{\bf w}_j^T{\bf H}_i{\bf f}_{i}{\bf x}_i^T
			+ 
			\sqrt{\rho}
			{\bf w}_j^T{\bf H}_{\textrm{SI},j}{\bf f}_{j}{\bf x}_j^T
			+
			{\bf w}_j^T{\bf Z}_{j}\\
			&=
			\underbrace{
			\sqrt{\rho}
			\tilde{h}_i{\bf x}_i^T}_{\textrm{Intended Signal}}
			+ 
			\underbrace{
			\sqrt{\rho}
			\tilde{h}_{\textrm{SI},j}{\bf x}_j^T}_{\textrm{Self-Interference}}
			+
			\underbrace{
			\tilde{\bf z}_{j}^T}_{\textrm{Noise}},
		\end{aligned}
	\end{equation}
	where 
	$\tilde{h}_i = {\bf w}_j^T{\bf H}_i{\bf f}_{i}$
	is the effective intended channel,
	${\tilde{h}_{\textrm{SI},j} \hspace{-0.5ex}=\hspace{-0.5ex} {\bf w}_j^T{\bf H}_{\textrm{SI},j}{\bf f}_{j}}$
	is the effective SI channel,
	$\tilde{\bf z}_{j} \hspace{-0.5ex}=\hspace{-0.5ex} {\bf w}_j^T{\bf Z}_{j}\hspace{-0.5ex}\in\hspace{-0.5ex}\mathbb{C}^{{Q}}$
	is the modified noise vector after the receive ABF.

\subsection{Structured Grassmannian Constellation}\label{secsub_CS}
We follow a particular structured Grassmannian constellation design, called cube-split \cite{cubeSplitConstellation}, where ${Q}$ Grassmannian cells are formed to construct the data symbol vector ${\bf x}\in \mathbb{C}^{{Q}}$.
	First, a single dimension of ${\bf x}$ is exploited for the selection of cells $\mathcal{C}_1, \cdots, \mathcal{C}_Q$.	
The Voronoi region associated with the cell $\mathcal{C}_q$ with $q=1,\cdots,Q$ is defined as:
\begin{equation}\label{eq_Cell_q}
	\mathcal{C}_q
	=
	\left\{
	{\bf x}=\left[x_1,\cdots,x_Q\right]^T:
	|x_q|>|x_p|, \forall p\ne q
	\right\}.
\end{equation}
The other $Q-1$ complex elements of ${\bf x}$ are converted to $2\left(Q-1\right)$ real coefficients to parameterize any points on a cell.
	Hence, $2\left(Q-1\right)$ real coefficients define an Euclidean space with $2\left(Q-1\right)$ dimensions. To form a grid on each cell, each dimension is  regularly divided  in the interval $\left[0,1\right]$.
When $B_v$ bits with $v=1,\cdots,2\left(Q-1\right)$ are used to characterize the $v^{th}$ dimension, the set $\mathcal{A}_v$ of uniformly spaced points maximizes the minimum distance  in the corresponding dimension:
\begin{equation}\label{eq_Ak}
	\mathcal{A}_v = \left\{
	\frac{1}{2^{B_v + 1}},
	\frac{3}{2^{B_v + 1}},
	\cdots,
	\frac{2^{B_v + 1} - 1}{2^{B_v + 1}}
	\right\},
\end{equation}
where 
	$2^{B_v}$ points are uniformly spread on the interval $\left[0,1\right]$.
The points in $\mathcal{A}_v$ are labeled via Gray encoding so that the neighboring	points differ by exactly one-bit.
	Then, a mapping function {generates} the  data symbol vector from a given cell having $2\left(Q-1\right)$ dimensional Euclidean space. 
	By using \eqref{eq_Cell_q} and \eqref{eq_Ak}, 
	the set of data symbol vectors is defined as:
\begin{equation}\label{eq_mapping}
	\mathcal{X}
	=\left\{
	{\bf x} 
	=
	r_q\left({\bf a}\right):~
	q=1,\cdots,Q, ~
	{\bf a}
	\in
	\mathop  \otimes \nolimits_{v=1}^{2\left(Q-1\right)}
	\mathcal{A}_v
	\right\},
\end{equation}
where $r_q\left(\cdot\right)$ is the mapping function, $\otimes_{v=1}^{2\left(Q-1\right)} \mathcal{A}_v$ represents the Cartesian product of the sets of $\mathcal{A}_v$ given in \eqref{eq_Ak}. 
It is worthwhile to remark that each element of 
${\bf a}\in\mathbb{R}^{2\left(Q-1\right)}$ 
is in the interval of $\left[0,1\right]$.
	Thus, the cube-split constellation contains 
$
Q2^{B_1+B_2+\cdots+B_{2\left(Q-1\right)}}$ possible data symbol vector ${\bf x}\in\mathbb{C}^Q$.
{If} the coherence block length $Q$ {is a} power of $2$, each data symbol vector generated via cube-split constellation carries:
\begin{equation}\label{eq_B}
	B
	=
	\log_2\left(Q\right)
	+
	\sum_{v=1}^{2\left(Q-1\right)}B_v ~\left[{\textrm{bits}}\right],
\end{equation}
bits of information in each coherence block. 
In order to express the mapping function given in \eqref{eq_mapping}, we first define
	$w_k
	=
	\frac{1}{\sqrt{2}}
	\mathcal{N}^{-1}\hspace{-0.5ex}\left(a_{2k-1}\right)
	+
	\frac{j}{\sqrt{2}}
	\mathcal{N}^{-1}\hspace{-0.5ex}\left(a_{2k}\right)
	\in \mathbb{C}
	$
and
	$t_k
	=
	\sqrt{
		\frac{
			1-\textrm{exp}\left(-{|w_k|^2}\right)
		}{
			1+\textrm{exp}\left(-{|w_k|^2}\right)
		}
	}
	\frac{w_k}{|w_k|} \in\mathbb{C}$ with $k=1,\cdots,Q-1$, 
where
	$\mathcal{N}\left(a\right)=\int_0^a {\frac{1}{{\sqrt {2\pi } }}{e^{ - {y^2}/2}}dy}$ 
is the cumulative distribution function (CDF) of standard  normal distribution and 
	$\mathcal{N}^{-1}\hspace{-0.5ex}\left(\cdot\right)$ is its inverse.
Afterwards, the mapping function is defined as:
\begin{equation}\label{eq_mapping_full}
	\begin{aligned}
		r_q\hspace{-0.25ex}\left(\hspace{-0.05ex}{\bf a}\hspace{-0.05ex}\right)
		\hspace{-0.5ex}=\hspace{-1.25ex}
		\sqrt{\hspace{-0.25ex}
			\frac{
				Q
			}{
				1\hspace{-0.5ex}+\hspace{-0.75ex}\sum\limits_{k=1}^{Q-1}
				\hspace{-0.5ex}\left|t_k\right|^2
		}}\hspace{-0.25ex}
		\left[t_1,\hspace{-0.25ex}\cdots\hspace{-0.25ex},t_{q-1},\hspace{-0.25ex}1\hspace{-0.05ex},t_q,\hspace{-0.25ex}\cdots\hspace{-0.35ex},t_{Q-1}\right]^T
		\hspace{-0.5ex}
		\in
		\hspace{-0.5ex}
		\mathbb{C}^Q,
	\end{aligned}
\end{equation}
where $|t_k|<1$ and the location of largest magnitude element depicts the selected Grassmannian cell as shown in \eqref{eq_Cell_q}.
Here,
	each data symbol vector satisfies the aforementioned power constraint 
		$
		\mathbb{E}\big\{||{\bf x}||^2\big\}
		=
		{Q}$.
	Also, the mapping function given in \eqref{eq_mapping_full} is bijective. Thus, for a given data symbol vector ${\bf x}$ with ${q} =
	\arg\max_p\left|x_p\right|$, the inverse mapping function is written as:
	\begin{equation}\label{eq_mapping_inv}
		\begin{aligned}
			{\bf a} = \hspace{2ex}&\hspace{-1.5ex} r_{{q}}^{-1}\left({\bf x}\right)\\
			\textrm{s.t. } 		
			&{\bf t}=\left[
			\frac{x_1}{x_{{q}}},
			\cdots,
			\frac{x_{{q}-1}}{x_{{q}}},
			\frac{x_{q+1}}{x_{{q}}},
			\cdots
			\frac{x_T}{x_{{q}}}
			\right]^T \in\mathbb{C}^{Q-1},\\
			&w_k=\sqrt{\log \frac{1+\left|t_k\right|^2}{1-\left|t_k\right|^2}}
			\frac{t_k}{\left|t_k\right|},
			\\
			&{a}_{2k-1}
			=\argmin_{a\in \mathcal{A}_{2k-1}}\left|a-\sqrt{2}\mathcal{N}\left(\textrm{Re}\left(w_k\right)\right)\right|,\\
			&{a}_{2k}
			=\argmin_{a\in \mathcal{A}_{2k}}\left|a-\sqrt{2}\mathcal{N}\left(\textrm{Im}\left(w_k\right)\right)\right|.
		\end{aligned}
	\end{equation}

In the proposed FD-NC mMIMO systems, the data symbol vectors ${\bf x}_1$ and ${\bf x}_2$ are generated via the cube-split constellation expressed in \eqref{eq_mapping} and \eqref{eq_mapping_full}. By using \eqref{eq_r_combined}, the combined signal at user $j$ can be simply decoded by applying the maximum likelihood (ML) decoder given as:
\begin{equation}
	\hat{\bf x}_{i,ML} = \argmax_{{\bf x}\in\mathcal{X}} \left| \tilde{\bf y}_{j}^T{\bf x}\right|^2, ~i,j\in\left\{1,2\right\}, j\ne i.
\end{equation}
To avoid high complexity ML decoder for the large constellation size, we employ a low-complexity greedy decoder \cite{cubeSplitConstellation}, where the inverse mapping function is applied to the left singular vector of the combined signal given in \eqref{eq_r_combined}. In other words, we first find $\hat{\bf u}=\argmax_{{\bf u}\in\mathbb{C}^Q, \left\|{\bf u} \right\|=1} |\tilde{\bf y}_j^T{\bf u}|$, then the cell index and position on the corresponding cell are decoded as $\hat{q} =	\arg\max_p\left|u_p\right|$
	and
$\hat{\bf a} = r_{\hat{q}}\left( \hat{\bf u}\right)$ by using \eqref{eq_mapping_inv}.
	Here, both ML and low-complexity greedy decoder are applied without a priori CSI for the non-coherent communications. 

	\begin{table}[t!]
	\caption{Cube-split constellation for $Q=2$, $B_1=B_2=1$.}
	\label{table1_example}
	\centering
	\renewcommand{\arraystretch}{1.75}
	\begin{tabular}{|c||c|c||c|c||c|}
		\hline
		\hspace{-1.5ex}
		\begin{tabular}[c]{@{}c@{}}Data\vspace{-2ex}\\ Bits\end{tabular} \hspace{-1.5ex} & 
		\hspace{-1ex}$\mathcal{C}_q$\hspace{-1.5ex} &
		${\bf a}^T$& 
		$w_1$ & 
		$t_1$ & 
		${\bf x}^T$ \\ \hline

		\hspace{-1ex}000\hspace{-1ex} &
		\hspace{-1ex}$\mathcal{C}_1$\hspace{-1.5ex} &
		\hspace{-1ex}$\left[\frac{1}{4}\hspace{-0.25ex},\hspace{-0.5ex}\frac{1}{4}\right]$\hspace{-1ex} &
		\hspace{-1ex}$-0.48 \hspace{-0.35ex}-\hspace{-0.35ex} j0.48$\hspace{-1ex} & 
		\hspace{-1ex}$-0.33 \hspace{-0.35ex}-\hspace{-0.35ex} j0.33$\hspace{-1ex} &
		\hspace{-1ex}$\left[1.28, -0.43 \hspace{-0.35ex}-\hspace{-0.35ex} j0.43\right]$ \hspace{-3.5ex}  ~ \\ \hline

		\hspace{-1ex}001\hspace{-1ex} & 
		\hspace{-1ex}$\mathcal{C}_1$\hspace{-1.5ex} &
		\hspace{-1ex}$\left[\frac{1}{4}\hspace{-0.25ex},\hspace{-0.5ex}\frac{3}{4}\right]$\hspace{-1ex} &
		\hspace{-1ex}$-0.48 \hspace{-0.35ex}+\hspace{-0.35ex} j0.48$\hspace{-1ex} & 
		\hspace{-1ex}$-0.33 \hspace{-0.35ex}+\hspace{-0.35ex} j0.33$\hspace{-1ex} &
		\hspace{-1ex}$\left[1.28, -0.43 \hspace{-0.35ex}+\hspace{-0.35ex} j0.43\right]$ \hspace{-3.5ex}  ~ \\ \hline
		
		\hspace{-1ex}010\hspace{-1ex} &
		\hspace{-1ex}$\mathcal{C}_1$\hspace{-1.5ex} &
		\hspace{-1ex}$\left[\frac{3}{4}\hspace{-0.25ex},\hspace{-0.5ex}\frac{1}{4}\right]$\hspace{-1ex} &
		\hspace{-1ex}$+0.48 \hspace{-0.35ex}-\hspace{-0.35ex} j0.48$\hspace{-1ex} & 
		\hspace{-1ex}$+0.33 \hspace{-0.35ex}-\hspace{-0.35ex} j0.33$\hspace{-1ex} &
		\hspace{-1ex}$\left[1.28, +0.43 \hspace{-0.35ex}-\hspace{-0.35ex} j0.43\right]$ \hspace{-3.5ex}  ~ \\ \hline
		
		\hspace{-1ex}011\hspace{-1ex} &
		\hspace{-1ex}$\mathcal{C}_1$\hspace{-1.5ex} &
		\hspace{-1ex}$\left[\frac{3}{4}\hspace{-0.25ex},\hspace{-0.5ex}\frac{3}{4}\right]$\hspace{-1ex} &
		\hspace{-1ex}$+0.48 \hspace{-0.35ex}+\hspace{-0.35ex} j0.48$\hspace{-1ex} & 
		\hspace{-1ex}$+0.33 \hspace{-0.35ex}+\hspace{-0.35ex} j0.33$\hspace{-1ex} &
		\hspace{-1ex}$\left[1.28, +0.43 \hspace{-0.35ex}+\hspace{-0.35ex} j0.43\right]$ \hspace{-3.5ex}  ~ \\ \hline

		\hspace{-1ex}100\hspace{-1ex} &
		\hspace{-1ex}$\mathcal{C}_2$\hspace{-1.5ex} &
		\hspace{-1ex}$\left[\frac{1}{4}\hspace{-0.25ex},\hspace{-0.5ex}\frac{1}{4}\right]$\hspace{-1ex} &
		\hspace{-1ex}$-0.48 \hspace{-0.35ex}-\hspace{-0.35ex} j0.48$\hspace{-1ex} & 
		\hspace{-1ex}$-0.33 \hspace{-0.35ex}-\hspace{-0.35ex} j0.33$\hspace{-1ex} &
		\hspace{-1ex}$\left[-0.43 \hspace{-0.35ex}-\hspace{-0.35ex} j0.43, 1.28\right]$ \hspace{-3.5ex}  ~ \\ \hline

		\hspace{-1ex}101\hspace{-1ex} &
		\hspace{-1ex}$\mathcal{C}_2$\hspace{-1.5ex} &
		\hspace{-1ex}$\left[\frac{1}{4}\hspace{-0.25ex},\hspace{-0.5ex}\frac{3}{4}\right]$\hspace{-1ex} &
		\hspace{-1ex}$-0.48 \hspace{-0.35ex}+\hspace{-0.35ex} j0.48$\hspace{-1ex} & 
		\hspace{-1ex}$-0.33 \hspace{-0.35ex}+\hspace{-0.35ex} j0.33$\hspace{-1ex} &
		\hspace{-1ex}$\left[-0.43 \hspace{-0.35ex}+\hspace{-0.35ex} j0.43, 1.28\right]$ \hspace{-3.5ex}  ~ \\ \hline
		
		\hspace{-1ex}110\hspace{-1ex} &
		\hspace{-1ex}$\mathcal{C}_2$\hspace{-1.5ex} &
		\hspace{-1ex}$\left[\frac{3}{4}\hspace{-0.25ex},\hspace{-0.5ex}\frac{1}{4}\right]$\hspace{-1ex} &
		\hspace{-1ex}$+0.48 \hspace{-0.35ex}-\hspace{-0.35ex} j0.48$\hspace{-1ex} & 
		\hspace{-1ex}$+0.33 \hspace{-0.35ex}-\hspace{-0.35ex} j0.33$\hspace{-1ex} &
		\hspace{-1ex}$\left[+0.43 \hspace{-0.35ex}-\hspace{-0.35ex} j0.43, 1.28\right]$ \hspace{-3.5ex}  ~ \\ \hline
		
		\hspace{-1ex}111\hspace{-1ex} & 
		\hspace{-1ex}$\mathcal{C}_2$\hspace{-1.5ex} &
		\hspace{-1ex}$\left[\frac{3}{4}\hspace{-0.25ex},\hspace{-0.5ex}\frac{3}{4}\right]$\hspace{-1ex} &
		\hspace{-1ex}$+0.48 \hspace{-0.35ex}+\hspace{-0.35ex} j0.48$\hspace{-1ex} & 
		\hspace{-1ex}$+0.33 \hspace{-0.35ex}+\hspace{-0.35ex} j0.33$\hspace{-1ex} &
		\hspace{-1ex}$\left[+0.43 \hspace{-0.35ex}+\hspace{-0.35ex} j0.43, 1.28\right]$ \hspace{-3.5ex}  ~ \\ \hline
	\end{tabular}
\end{table}

Table \ref{table1_example} presents {the constellation points} ${\bf x}\in\mathbb{C}^2$ based on the cube-split constellation for $Q=2$ and $B_1=B_2=1$, where
	$B=\log_2\left(Q\right)+B_1+B_2=3$ bits are transmitted in a coherence block.
Here, the first data bit selects a Grassmannian cell $\mathcal{C}_1$ or $\mathcal{C}_2$, then the last two data bits determine the position on corresponding cell.
	For instance, when the first data bit is $0$, the first cell $\mathcal{C}_1$ is chosen and we have $|x_1|>|x_2|$. Otherwise, $|x_2|>|x_1|$ is valid if the first data bit is $1$.
On the other hand, there are $2\left(Q-1\right)=2$ dimensions in each Grassmannian cell, where each dimension is characterized via $B_1=B_2=1$ bit with $\mathcal{A}_1=\mathcal{A}_2=\left\{\frac{1}{4},\frac{3}{4}\right\}$ as shown in \eqref{eq_Ak}. 
	Hence, the last two bits are used to construct ${\bf a}=\left[a_1,a_2\right]^T$. 
Specifically, as shown from Table \ref{table1_example}, the second and third data bits correspond to $a_1$ and $a_2$, respectively.
	For a given $a_1$ and $a_2$, one can calculate 
$w_1
=
\frac{1}{\sqrt{2}}
\mathcal{N}^{-1}\hspace{-0.5ex}\left(a_{1}\right)
+
\frac{j}{\sqrt{2}}
\mathcal{N}^{-1}\hspace{-0.5ex}\left(a_{2}\right)
$
and
$t_1
=
\sqrt{
	\frac{
		1-\textrm{exp}\left(-{|w_1|^2}\right)
	}{
		1+\textrm{exp}\left(-{|w_1|^2}\right)
	}
}
\frac{w_1}{|w_1|}$.
	Finally, each data symbol vector ${\bf x}$ is calculated by substituting $t_1$ into \eqref{eq_mapping} and \eqref{eq_mapping_full}.

\section{Channel Model}\label{sec_Channel}
Based on the 3D geometry-based mmWave channel model and ULA structure, the channel matrix from user $i$ to user $j$ is defined as follows \cite{ASIL_ABHP_Access,Satyanarayana2019,ASIL_FD_SIC,Report_5G_UMi_UMa_Rel16}:
	\begin{equation}\label{eq_H_i}
		\begin{aligned}
			{\bf H}_i 
			=
			\sum_{l=1}^{L}
			\frac{\beta_{i,l}}{
			\sqrt{\alpha \tau_{i,l}^{\eta}}}
			\bm{\phi}_{r,i}  \left(\varphi_{j,l}\right)
			\bm{\phi}_{t,i}^H\left(\theta_{i,l}\right)
			=
			\bm{\Phi}_{r,j} {\bf G}_i \bm{\Phi}_{t,i},
		\end{aligned}
	\end{equation}
	where 
	$L$
	is the total number of paths,
	$\beta_{i,l}\sim\mathcal{CN}\big(0,\frac{1}{L}\big)$
	and
	$\tau_{i,l}$
	are respectively the path gain and distance for the $l^{th}$ path,
	$\eta$
	is the path loss exponent.
	The reference path loss is defined as $\alpha = 32.4 + 20\log_{10}\left(f_c\right)$ in dB, where the carrier frequency $f_c$ is in GHz  \hspace{-0.2ex}\cite{Report_5G_UMi_UMa_Rel16}.
	Hence, 	${\bf G}_i
	\hspace{-0.5ex}=\hspace{-0.5ex}
	\frac{1}{\sqrt{\alpha}}
	\textrm{diag}
	\Big(
	\hspace{-0.25ex}
	\frac{\beta_{i,1}}{\sqrt{\tau_{i,1}^{\eta}}},
	\hspace{-.5ex}\cdots\hspace{-0.5ex},
	\frac{\beta_{i,L}}{\sqrt{\tau_{i,L}^{\eta}}}
	\hspace{-0.25ex}
	\Big)
	\hspace{-0.5ex}\in
	\hspace{-0.5ex}\mathbb{C}^{L\times L}$	
	is the diagonal path gain matrix.
		Moreover,
	$\bm{\Phi}_{t,i}\in\mathbb{C}^{L \times M_i}$
	is the transmit phase response matrix with the rows of:
	\begin{equation}\label{eq_tx_PhaseResponse}
		\bm{\phi}_{t,i}\hspace{-0.5ex}\left(\theta_{i,l}\right)
		\hspace{-0.5ex}=\hspace{-1ex}
		\left[\hspace{-0.25ex}
		1,\hspace{-0.25ex}
		e^{-j2\pi d \cos\left(\theta_{i,l}\right)}\hspace{-0.5ex},
		\hspace{-0.5ex}\cdots\hspace{-0.5ex},\hspace{-0.25ex}
		e^{-j2\pi d \left(M_i-1\right)\cos\left(\theta_{i,l}\right)}\hspace{-0.25ex}
		\right]\hspace{-0.5ex},
	\end{equation}
	where 
	$d=0.5$ is the normalized half-wavelength distance between antennas,
	$\theta_{i,l}
	\in
	\left[
	\theta_i-\delta_i^\theta,
	\theta_i+\delta_i^\theta
	\right]$
	is 
	AoD at user $i$ for the $l^{th}$ path
	with mean $\theta_i$ and spread $\delta_i^\theta$.
	Then,
	$\bm{\Phi}_{r,j}\in\mathbb{C}^{N_j \times L}$
	is the receive phase response matrix with the columns of:
	\begin{equation}\label{eq_rx_PhaseResponse}
		\bm{\phi}_{r,j}\hspace{-0.5ex}\left(\varphi_{j,l}\right)
		\hspace{-0.5ex}=\hspace{-1ex}
		\left[\hspace{-0.25ex}
		1,
		e^{j2\pi d \cos\left(\varphi_{j,l}\right)},
		\hspace{-0.5ex}\cdots\hspace{-0.5ex},\hspace{-0.25ex}
		e^{j2\pi d \left(N_j-1\right)\cos\left(\varphi_{j,l}\right)}
		\hspace{-0.25ex}\right]^{\hspace{-0.35ex}T}\hspace{-0.5ex},
	\end{equation}
	where
	$\varphi_{j,l}
	\hspace{0.5ex}
	\in
	\left[
	\varphi_j-\delta_j^\varphi,
	\varphi_j+\delta_j^\varphi
	\right]$
	is AoA at user $j$ for the $l^{th}$ path
	with mean $\varphi_j$ and spread $\delta_j^\varphi$.
	
\begin{figure}[!t]
		\centering
		\includegraphics[width = 0.45\columnwidth]{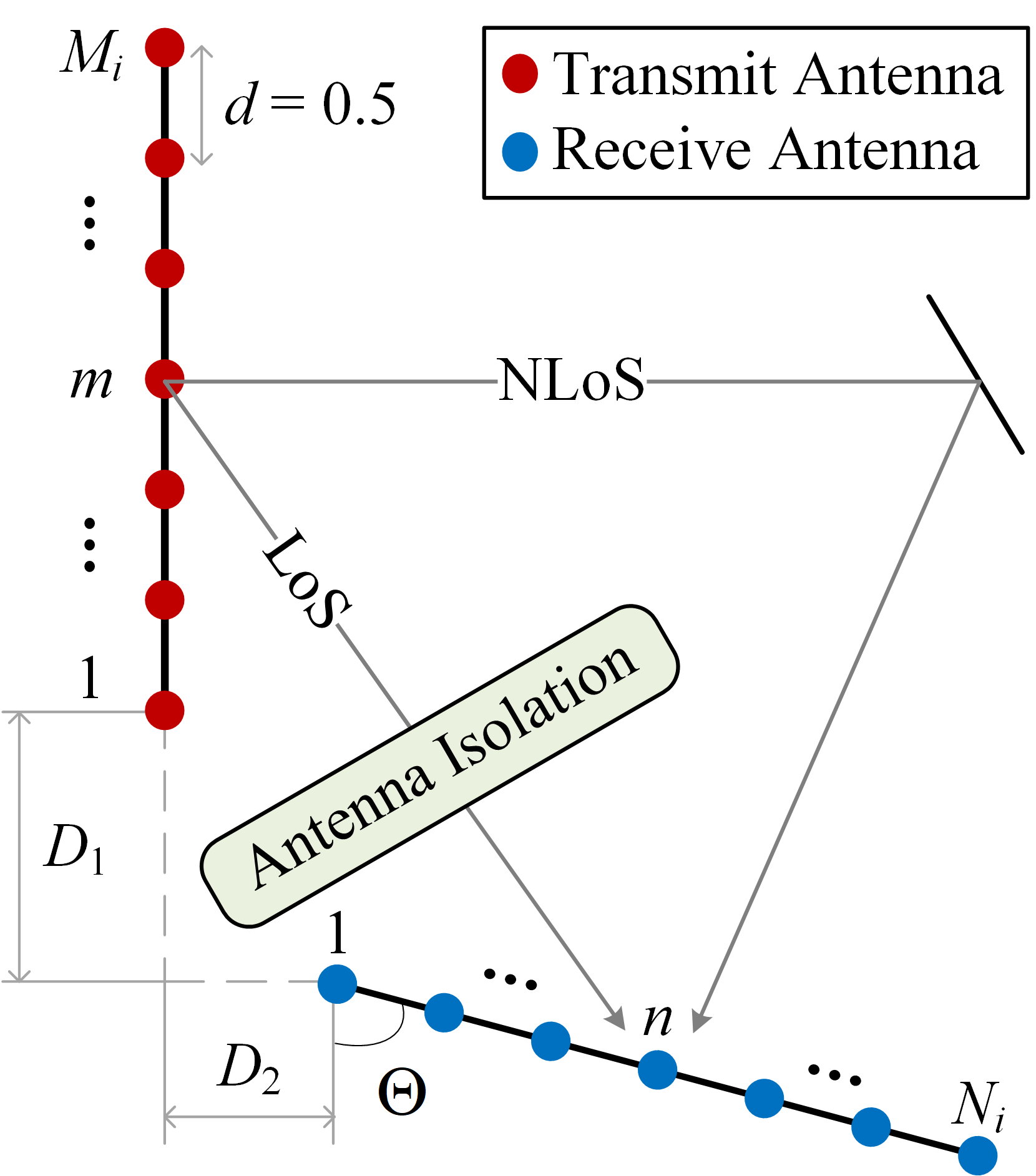}
		\caption{Transmit and receive antenna array configuration.}
		\label{fig3_TxRxArray}
		\vspace{-3ex}
	\end{figure}

	As illustrated in Fig. \ref{fig3_TxRxArray}, the SI channel includes the residual near-field SI channel with the line-of-sight (LoS) paths and the far-field SI channel with the non-line-of-sight (NLoS) paths \cite{Satyanarayana2019,ASIL_FD_SIC}.
We denote by 
	$D_1$ ($D_2$)  the vertical (horizontal) distance between the transmit and receive ULAs normalized by the wavelength, and $\Theta$ the rotation angle of the receive ULA with respect to the transmit ULA.
		The antenna isolation block reduces the SI channel power, especially the strong near-field components. 
	Thus, the complete SI channel matrix at user $i$ is modeled as follows:
	\begin{equation}\label{eq_H_SI}
		{\bf H}_{\textrm{SI},i}
		=
		{\bf H}_{\textrm{LoS},i}
		+
		{\bf H}_{\textrm{NLoS},i}\in\mathbb{C}^{N_i\times M_i},
	\end{equation}
	where
	${\bf H}_{\textrm{LoS},i}\in\mathbb{C}^{N_i \times M_i}$
	is the residual near-field SI channel,
	${\bf H}_{\textrm{NLoS},i}\in\mathbb{C}^{N_i \times M_i}$
	is the far-field SI channel.
	Based on the spherical wavefront\footnote{Because of the short distance between transmit/receive ULAs, the spherical wavefront is more realistic than the planar wavefront for ${\bf H}_{\textrm{LoS},i}$ \cite{Satyanarayana2019,ASIL_FD_SIC}.}, we define the near-field SI channel as:
	\begin{equation}
		{\bf H}_{\textrm{LoS},i}\left[n,m\right]=\frac{\kappa}{\Delta_{m \rightarrow n}}e^{-j2\pi \Delta_{m \rightarrow n}},
	\end{equation}
	where 
	$~m=1,\cdots,M_i$ and $n=1,\cdots,N_i$ are the index of transmit and receive antennas, respectively,
	$\Delta_{m \rightarrow n}$ is the distance between the corresponding antennas,
	$\kappa$ is the normalization scalar to satisfy
	$10\log_{10}\big(\mathbb{E}\left\| {\bf H}_{\textrm{LoS},i}\right\|_F^2\big)=-P_{\textrm{IS,dB}}$
	as the residual near-field SI channel power. Here, $P_{\textrm{IS,dB}}$ is the amount of SIC achieved by the antenna isolation\footnote{When there is no antenna isolation, we have $P_{\textrm{IS,dB}}=0$ dB and the unity power near-field SI channel, i.e.,  $\mathbb{E}\big\{\|{{\bf{H}}_{\textrm{LoS},i}}\|_F^2\big\}=1$. On the other hand, when the antenna isolation block achieves $P_{\textrm{IS,dB}}=20$ dB SIC, the average residual near-field SI channel power is measured as $\mathbb{E}\big\{\|{{\bf{H}}_{\textrm{LoS},i}}\|_F^2\big\}=0.01$.}.
	According to the array configuration in Fig. \ref{fig3_TxRxArray}, the distance between $m^{th}$ transmit and $n^{th}$ receive antenna elements is calculated as:
	\begin{equation}
		\begin{aligned}
			\Delta_{m \rightarrow n}
			&=
			\left(\left[D_1
			+
			\left(m-1\right)d
			+
			\cos\left(\Theta\right)\left(n-1\right)d\right]^2\right.\\
			&+
			\left.\left[D_2
			+
			\sin\left(\Theta\right)\left(n
			+
			1\right)d\right]^2\right)^{\frac{1}{2}}.
		\end{aligned}
	\end{equation}
	On the other hand, we consider the 3D geometry-based mmWave channel model for the far-field SI channel 
	${\bf H}_{\textrm{NLoS},i}$
	since the planar wavefront is applicable for the NLoS paths.
	Similar to \eqref{eq_H_i}, the far-field SI channel is modeled as:
	\begin{equation}\label{eq_H_NLoS}
		\begin{aligned}
			{\bf H}_{\textrm{NLoS},i}
			=
			\bm{\Phi}_{\textrm{SI},r,i} {\bf G}_{\textrm{SI},i} \bm{\Phi}_{\textrm{SI},t,i},
		\end{aligned}
	\end{equation}
	where 
	${\bf G}_{\textrm{SI},i}
	=
	\frac{1}{\sqrt{\alpha}}
	\textrm{diag}
	\Big(
	\frac{\beta_{\textrm{SI},i,1}}{\sqrt{\tau_{\textrm{SI},i,1}^{\eta}}},
	\cdots,
	\frac{\beta_{\textrm{SI},i,L_\textrm{SI}}}{\sqrt{\tau_{\textrm{SI},i,L_\textrm{SI}}^{\eta}}}
	\Big)
	\in\mathbb{C}^{L_{\textrm{SI}}\times L_{\textrm{SI}}}$
	with 
	$L_{\textrm{SI}}$ paths,
	$\bm{\Phi}_{\textrm{SI},r,i}\in\mathbb{C}^{N_i\times L_\textrm{SI}}$ 
	and 
	$\bm{\Phi}_{\textrm{SI},t,i}\in\mathbb{C}^{L_\textrm{SI}\times M_i}$
	are the receive and transmit phase response matrices, respectively.
	Here, we consider the mean AoA (AoD)
	$\varphi_{\textrm{SI},i}$ ($\theta_{\textrm{SI},i}$)
	and the AoA (AoD) spread
	$\delta_{\textrm{SI},i}^{\varphi}$ ($\delta_{\textrm{SI},i}^{\theta}$)
	to build the corresponding phase response matrices.

\section{Analog Beamformer Design}\label{sec_ABF}
Throughout this section, we develop the transmit/receive analog beamformer (ABF) for the FD-NC mMIMO systems.
	Here, we have three primary objectives: 
		(i)		design ABF without instantaneous CSI for non-coherent communications,
		(ii) 	maximize the intended signal power, 
		(iii) 	suppress the SI power.
Regarding the first objective, the transmit/receive ABF is developed via the slow time-varying AoD/AoA parameters\footnote{As shown in \cite{ASIL_Xiaoyi_DL_CE}, a geospatial data-based offline estimation technique can be utilized to  efficiently acquire the mean AoD/AoA and their spread instead of applying the traditional online channel sounding.}.
	
By using \eqref{eq_r_combined} and \eqref{eq_H_i}, the effective channel is rewritten as:
\begin{equation}
	\begin{aligned}
		\tilde{h}_i 
		&= 
		{\bf w}_j^T{\bf H}_i{\bf f}_{i}=
		{\bf w}_j^T\bm{\Phi}_{r,j} {\bf G}_i \bm{\Phi}_{t,i}{\bf f}_{i}.
	\end{aligned}
\end{equation}
Regarding the second objective, the transmit (receive) ABF should belong to the subspace spanned by $\bm{\Phi}_{t,i}$ ($\bm{\Phi}_{r,j}$) for maximizing the intended signal power given in \eqref{eq_r_combined}, i.e., 
	$\textrm{Span}\left({\bf f}_i\right)\subset 
	\textrm{Span}\left(\bm{\Phi}_{t,i}\right)$ 
and 
	$\textrm{Span}\left({\bf w}_j\right)\subset
	\textrm{Span}\left(\bm{\Phi}_{r,j}\right)$.
From \eqref{eq_H_i}, \eqref{eq_tx_PhaseResponse} and \eqref{eq_rx_PhaseResponse}, we define AoD and AoA supports for the intended channel as:
\begin{equation}\label{eq_AoD_Support_H_i}
	\begin{aligned}
		\textrm{AoD}_i
		&=
		\left\{
		\cos\left(\theta\right)
		\left|\right.
		\theta\in\left[
		\theta_i-\delta_i^\theta,
		\theta_i+\delta_i^\theta
		\right]
		\right\},\\
		\textrm{AoA}_j
		&=
		\left\{
		\cos\left(\varphi\right)
		\left|\right.
		\varphi\in\left[
		\varphi_j-\delta_j^\varphi,
		\varphi_j+\delta_j^\varphi
		\right]
		\right\}.
	\end{aligned}
\end{equation}
On the other hand, by combining \eqref{eq_r_combined}, \eqref{eq_H_SI} and \eqref{eq_H_NLoS}, we define an approximate zero condition for the effective SI channel as:
\begin{equation}\label{eq_approx_zero}
	\begin{aligned}
		\tilde{h}_{\textrm{SI},j} 
		&= 
		{\bf w}_j^T{\bf H}_{\textrm{SI},j}{\bf f}_{j}
		=
		{\bf w}_j^T
		\left(
		{\bf H}_{\textrm{LoS},j}
		+
		{\bf H}_{\textrm{NLoS},j}
		\right)
		{\bf f}_{j}\\
		&=
		{{\bf w}_j^T\bm{\Phi}_{\textrm{SI},r,j}}
		{\bf G}_{\textrm{SI},j} 
		{\bm{\Phi}_{\textrm{SI},t,j}{\bf f}_{j}}
		+
		{{\bf w}_j^T{\bf H}_{\textrm{LoS},j}{\bf f}_{j}}
		\approx 0.
	\end{aligned}
\end{equation}
After suppressing the near-field LoS part of SI channel via antenna isolation, the far-field NLoS part of SI channel become dominant \cite{FD_AntennaSep_74dB}. Thus, regarding the third objective, the transmit (receive) ABF should be in the null space of $\bm{\Phi}_{\textrm{SI},t,i}$ ($\bm{\Phi}_{\textrm{SI},r,j}$) 
for suppressing the SI power,
	i.e., $\textrm{Span}\left({\bf f}_j\right)\subset 
	\textrm{Null}\left(\bm{\Phi}_{\textrm{SI},t,j}\right)$ 
	and 
	$\textrm{Span}\left({\bf w}_j\right)\subset
	\textrm{Null}\left(\bm{\Phi}_{\textrm{SI},r,j}\right)$.
Similar to \eqref{eq_AoD_Support_H_i}, the AoD and AoA supports for the SI channel are respectively defined as:
\begin{equation}
	\begin{aligned}
		\textrm{AoD}_{\textrm{SI},j}
		&=
		\big\{
		\cos\left(\theta\right)
		\big|\big.
		\theta\in\big[
		\theta_{\textrm{SI},j}-\delta_{\textrm{SI},j}^\theta,
		\theta_{\textrm{SI},j}+\delta_{\textrm{SI},j}^\theta
		\big]
		\big\},\\
		\textrm{AoA}_{\textrm{SI},j}
		&=
		\big\{
		\cos\left(\varphi\right)
		\big|\big.
		\varphi\in\big[
		\varphi_{\textrm{SI},j}-\delta_{\textrm{SI},j}^\varphi,
		\varphi_{\textrm{SI},j}+\delta_{\textrm{SI},j}^\varphi
		\big]
		\big\}.
	\end{aligned}
\end{equation}
To cover the complete AoD and AoA supports, we define the quantized angles as
$\lambda_{m,i}=-1+\frac{2m-1}{M_i}$ with $m=1,\cdots,M_i$
and 
$\gamma_{n,i}=-1+\frac{2n-1}{N_i}$ with $n=1,\cdots,N_i$.
Also, by using \eqref{eq_tx_PhaseResponse} and \eqref{eq_rx_PhaseResponse}, the transmit and receive steering vectors at user $i$ are constructed as 
${\bf e}_{t,i}\left(\lambda_{m,i}\right)
=
\frac{1}{\sqrt{M_i}}
\bm{\phi}_{t,i}^H\left(\cos^{-1}\left(\lambda_{m,i}\right)\right)$
and
${\bf e}_{r,i}\left(\gamma_{n,i}\right)
=
\frac{1}{\sqrt{N_i}}
\bm{\phi}_{r,i}^*\left(\cos^{-1}\left(\gamma_{n,i}\right)\right)$,
respectively.
	For maximizing the intended signal power and suppressing the far-field SI channel power, we first derive the optimal quantized angle for the transmit ABF vector at user $i$ as:
\begin{equation}
	\begin{aligned}
		\lambda_{i}^{\textrm{opt}} \hspace{3.5ex}&\hspace{-4ex}= \argmax_{
			\substack{
				\lambda_{m,i}
			}
		}
		\left\|\bm{\Phi}_{t,i}{\bf e}_{t,i}\left(\lambda_{m,i}\right)\right\|_2^2\\
		\textrm{s.t. }
		& \lambda_{m,i}\hspace{-.25ex}\in\hspace{-.25ex}\textrm{AoD}_i, ~\lambda_{m,i}\hspace{-.25ex}\notin\hspace{-.25ex}\textrm{AoD}_{\textrm{SI},i},
		~m=1,\cdots,M_i.
	\end{aligned}
\end{equation}
Similarly, the optimal quantized angle for the receive ABF vector at user $j$ is obtained as:
\begin{equation}
	\begin{aligned}
		\gamma_{j}^{\textrm{opt}} \hspace{4ex}&\hspace{-4ex}= \argmax_{
			\substack{
				\gamma_{n,j}
			}
		}
		\left\|
		{\bf e}_{r,j}^T\left(\gamma_{n,j}\right)\bm{\Phi}_{r,j}
		\right\|_2^2\\
		\textrm{s.t. }
		& \gamma_{n,j}\in\textrm{AoA}_i, ~\gamma_{n,j}\notin\textrm{AoA}_{\textrm{SI},j},
		~n=1,\cdots,N_j.
	\end{aligned}
\end{equation}
Finally, we find the transmit and receive ABF vectors as:
\begin{equation}
	{\bf f}_i = {\bf e}_{t,i}\left(\lambda_i^{\textrm{opt}}\right),~~
	{\bf w}_j = {\bf e}_{r,j}\left(\gamma_j^{\textrm{opt}}\right).
\end{equation}

\section{Illustrative Results}\label{sec_RESULTS}

This section presents Monte Carlo simulation results for evaluating bit error rate (BER) of the proposed full-duplex non-coherent (FD-NC) mMIMO systems. We also compare its performance with other reference transmission schemes displayed in Fig. \ref{fig2_TransmissionSchemes}.
Table \ref{table2_sim}
\begin{table}[t!]
	\caption{Simulation parameters.}
	\vspace{-1ex}
	\label{table2_sim}
	\centering
	\renewcommand{\arraystretch}{1.2}
	\begin{tabular}{|l|l|}
		\hline
		{Antenna Array Size} & $M=N=64$      \\ \hline
		{Coherence Block Length} & $Q\in\left\{2,4\right\}$\\ \hline
		{Average Bits per Coherence Block} & $B\in\left\{3,5,8,14\right\}$\\ \hline
		{Carrier Frequency \cite{Report_5G_UMi_UMa_Rel16}} & $f_c=28$ GHz\\ \hline
		{Bandwidth \cite{Report_5G_UMi_UMa_Rel16}} & $100$ MHz\\ \hline
		{Reference Path loss \cite{Report_5G_UMi_UMa_Rel16}} & $\alpha= 61.34$ dB\\ \hline
		{Path Loss Exponent}\cite{Report_5G_UMi_UMa_Rel16} & $\eta=2.1$ \\ \hline
		{Noise PSD \cite{ASIL_FD_SIC}} & $-174$ dBm/Hz      \\ \hline
		{Number of paths\cite{Report_5G_UMi_UMa_Rel16}} & $L=L_{\textrm{SI}}=20$      \\ \hline
		Transmit/Receive Array Configuration & $D_1=2$, $D_2=0$, $\Theta=0$\\ \hline
		Number of Network Realizations  & $10^8$\\ \hline\hline
		${\bf H}_i$: Mean AoD/AoA & $\theta_i=105^\circ$, $\varphi_i=65^\circ$\\ \hline
		${\bf H}_i$: AoD/AoA Spread \cite{Report_5G_UMi_UMa_Rel16} & $\delta^\theta_i=\delta^\varphi_i=5^\circ$\\ \hline\hline
		${\bf H}_{\textrm{SI},i}$: Path Distance & ${\tau_{\textrm{SI},i}\in\left[5,15\right]}$\\ \hline
		${\bf H}_{\textrm{SI},i}$: Mean AoD/AoA & $\theta_{\textrm{SI},i}=20^\circ$, $\varphi_{\textrm{SI},i}=160^\circ$\\ \hline
		${\bf H}_{\textrm{SI},i}$: AoD/AoA Spread \cite{Report_5G_UMi_UMa_Rel16} & $\delta^\theta_{\textrm{SI},i} =\delta^\varphi_{\textrm{SI},i} =5^\circ$\\ \hline
	\end{tabular}
	\vspace{-2.2ex}
\end{table}
outlines the simulation setup according to the recent 3GPP Release 16 specifications \cite{Report_5G_UMi_UMa_Rel16}.
It is worthwhile to remark that $B$ denotes the average number of bits transmitted by each user per coherence block.

\begin{figure}[!t]
	\centering
	\includegraphics[width = \columnwidth]{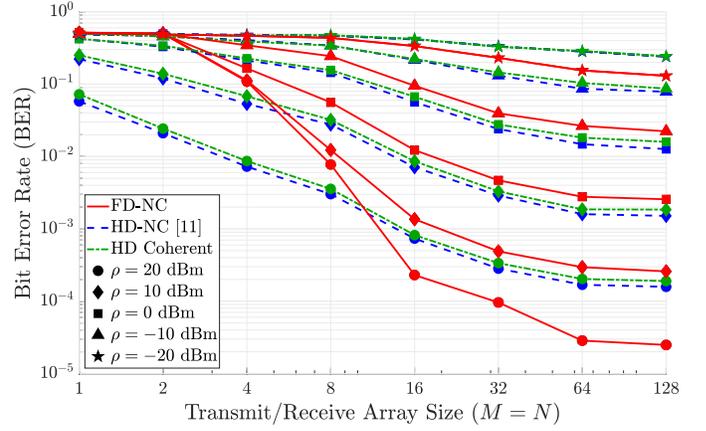}
	\vspace{-3ex}
	\caption{BER performance versus array size ($Q\hspace{-.5ex}=\hspace{-.5ex}2$, $B\hspace{-.5ex}=\hspace{-.5ex}3$ bits).}
	\label{fig4_BER_vs_M}
	\vspace{-2ex}
\end{figure}

Fig. \ref{fig4_BER_vs_M} shows the BER of FD-NC, HD-NC\footnote{Albeit HD-NC is proposed for single-input multiple output (SIMO) systems in \cite{cubeSplitConstellation}, we here adopt it for mMIMO via the proposed ABF design.} \cite{cubeSplitConstellation} and HD coherent transmission schemes versus transmit/receive {ULA} size, where each user transmits $B=3$ bits on average per coherence block length of $Q=2$. Here, we consider $P_{\textrm{IS,dB}} =74$ dB antenna isolation as in \cite{FD_AntennaSep_74dB}.
	FD-NC requires only $B_1=B_2=1$ bit per each dimension in the cube-split constellation according to \eqref{eq_B}.
In the HD-NC scheme, the same average transmission rate of $B=3$ bits is achieved, when one HD user transmits $2.5$ bits per coherence block via $B_1=B_2=2$, while other HD user sends $3.5$ bits per coherence block via $B_1=B_2=3$. 
	For the HD coherent transmission with $Q=2$, the first symbol interval is exploited for the pilot-based channel estimation, then the data signal is transmitted in the second symbol interval.
Thus, each user should use $64$-QAM to ensure the same average transmission rate of $B=3$ bits per coherence block.
	Also, the coherent receivers apply minimum mean square error (MMSE) estimator\footnote{According to \eqref{eq_r_combined} and Fig. \ref{fig2_TransmissionSchemes}, the received pilot signal after receive ABF is written as $\tilde{y}_j=\sqrt{\rho} \tilde{h}_i+z_j$ with $z_j\sim\mathcal{CN}\left(0,\sigma_z^2\right)$. 
		Then, the MMSE channel estimation is obtained as $\hat{h}_i=\frac{\sqrt{\rho}}{\sigma_z^2+\rho}\tilde{y}_j$ \cite{ASIL_FD_SM_I_CSI_ICT}.} to estimate the effective intended channel $\tilde{h}_i$ given in \eqref{eq_r_combined}.
Numerical results show that even though the antenna isolation provides $P_{\textrm{IS,dB}} =74$ dB SIC for the near-field SI signal, FD-NC with small transmit/receive antenna arrays severely suffers from the dominant far-field SI signal. For example, when we have a {single-input single-output (SISO) system}, i.e., $M=N=1$, FD-NC has the BER value of $0.5$ independent from transmit power $\rho$.
The main reason is the lack of beamforming capability, which prevents the cancellation of dominant far-field SI channel (please see \eqref{eq_approx_zero}).
	{As} the array size increases, the proposed ABF technique further enhances the SIC quality, which reduces the detection error  of FD-NC.
Thus, the error performance of FD-NC improves for larger arrays  {in mMIMO systems. Then, it} outperforms both of its HD counterparts.
		To illustrate, FD-NC has lower BER values in every transmit power scenarios for $M=N\ge 16$.
Also, HD-NC achieves better error performance than the HD coherent scheme in every transmit power and array size scenarios.
 
\begin{figure}[!t]
	\begin{subfigure}{0.48\columnwidth}
		\centering
		\includegraphics[width = \textwidth]{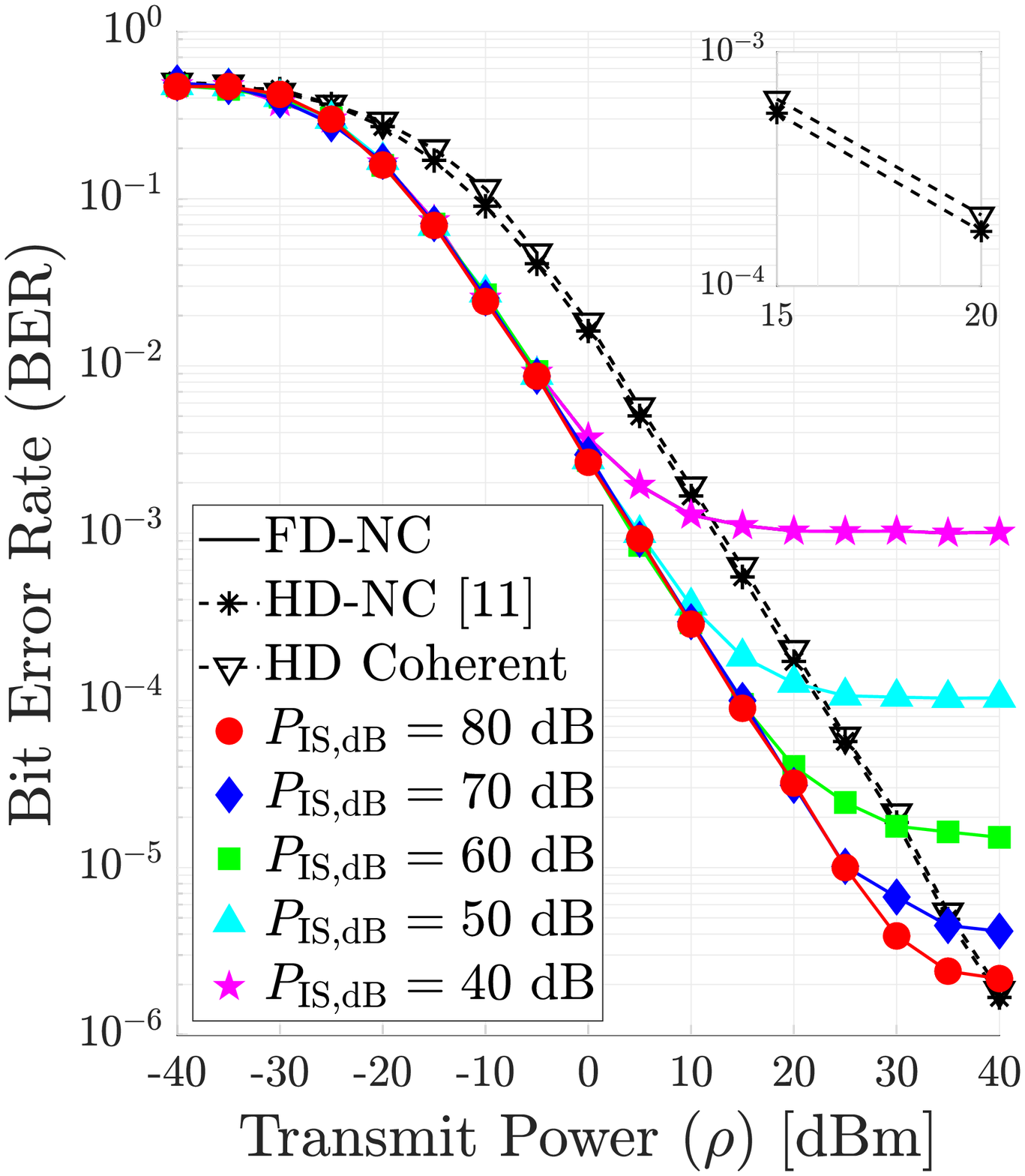}
		\caption{$B=3$ bits}
	\end{subfigure}
	\hfill
	\begin{subfigure}{0.48\columnwidth}
		\centering
		\includegraphics[width = \textwidth]{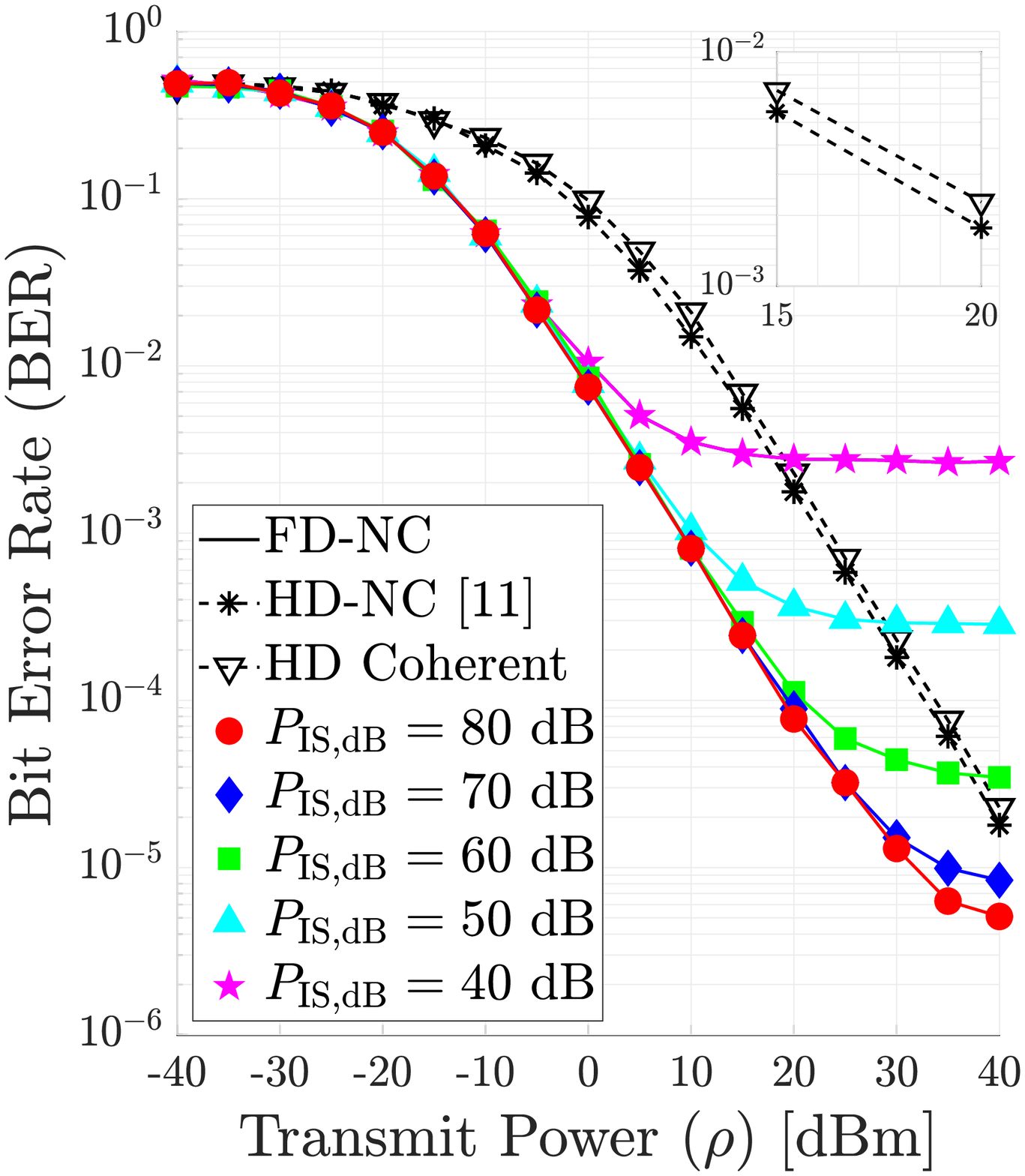}
		\caption{$B=5$ bits}
	\end{subfigure}
	\vspace{-0.5ex}
	\caption{BER performance versus transmit power ($Q=2$).}
	\vspace{-2ex}
	\label{fig5}
\end{figure}

In Fig. \ref{fig5}, we investigate the effect of antenna isolation on FD-NC via varying $P_{\textrm{IS,dB}}$ between $40$ dB and $80$ dB, where each user has $M=64$ transmit and $N=64$ receive antennas. 
For coherence block of $Q=2$ symbol intervals, Fig. \ref{fig5}(a) and Fig. \ref{fig5}(b) illustrate the BER curves versus the transmit power for $B=3$ and $B=5$ bits, respectively.
{In Fig. \ref{fig5}(a) for $B=3$ bits, we observe a $7.6$ dB performance gap between FD-NC and HD-NC at the BER value of $10^{-3}$. 
	Furthermore, when the transmission rate increases to $B=5$ bits in Fig. \ref{fig5}(b), the performance gap is greatly enlarged to $13.5$ dB for the benefit of FD-NC.
	Because the simultaneous FD transmission requires a smaller constellation size compared to its HD counterparts.
	Regarding the HD transmission with $B=3$ bits ($B=5$ bits), the non-coherent scheme requires $0.5$ dB ($1.1$ dB) less transmit power to achieve the same BER value in the coherent scheme.}
	It is seen that FD-NC converges to an error floor as the transmit power increases due to the residual SI. The error floors can be lowered based on the antenna isolation capability.
For instance, the error floor for $B=3$  bits can be reduced from $10^{-4}$ to $4\times10^{-6}$ by enhancing the antenna isolation from $P_{\textrm{IS,dB}}=50$ dB to $P_{\textrm{IS,dB}}=70$ dB. 
	Moreover, when the antenna isolation increases, the error floor is monitored in the higher transmit power regime, {where we obtain lower BER}. 

\begin{figure}[!t]
	\begin{subfigure}{0.48\columnwidth}
		\centering
		\includegraphics[width = \textwidth]{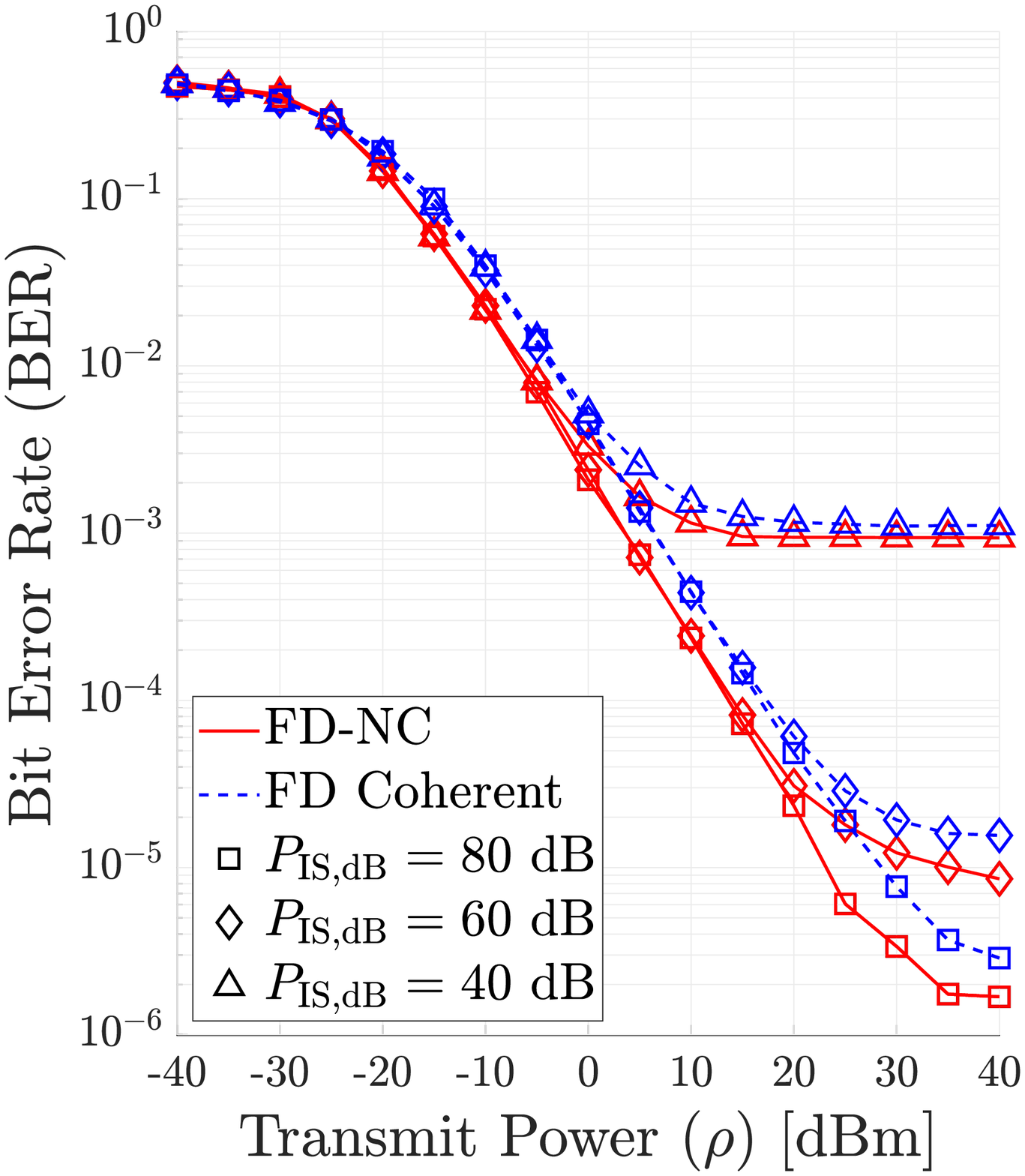}
		\caption{$B=8$ bits}
	\end{subfigure}
	\hfill
	\begin{subfigure}{0.48\columnwidth}
		\centering
		\includegraphics[width = \textwidth]{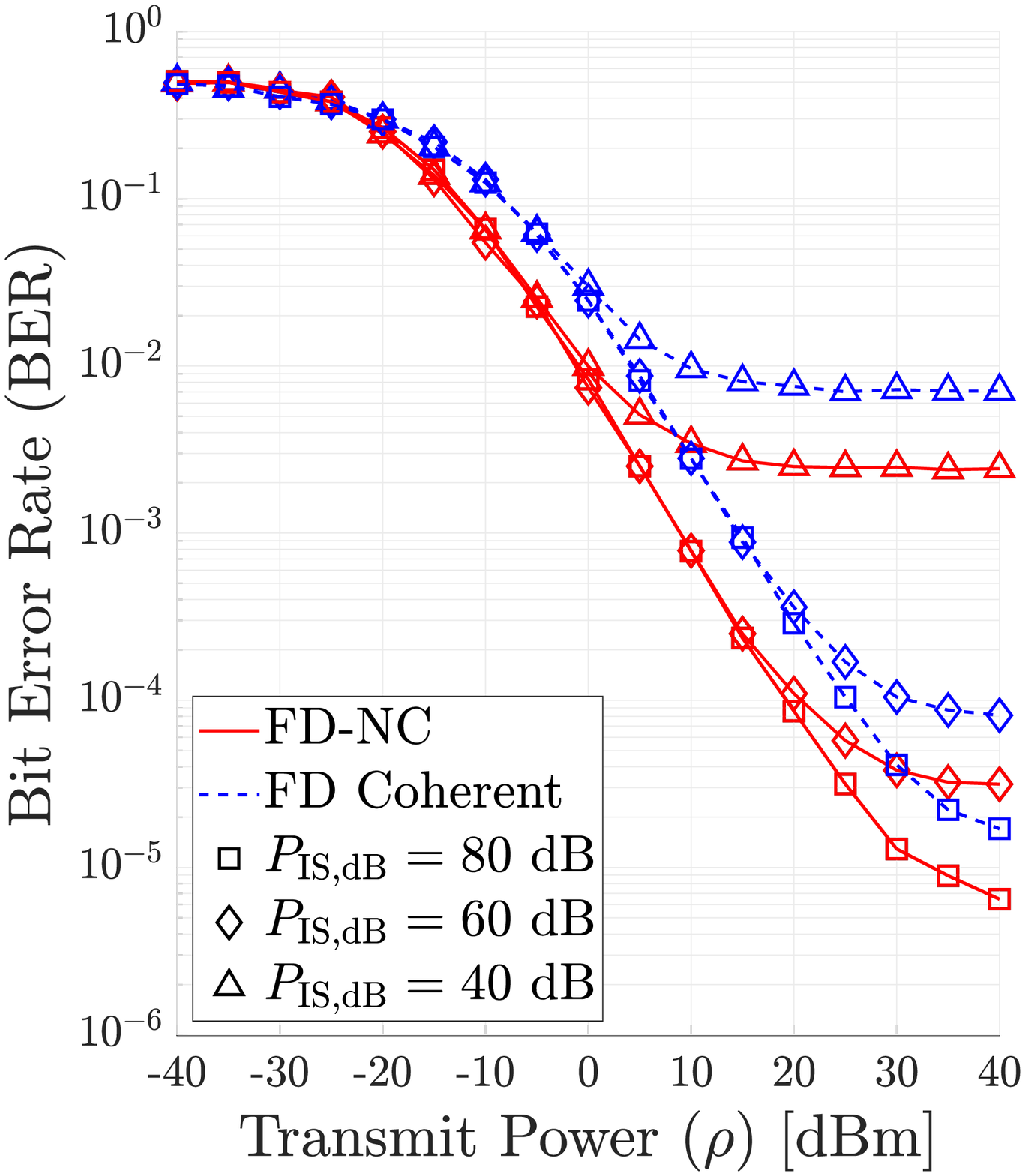}
		\caption{$B=14$ bits}
	\end{subfigure}
	\vspace{-0.5ex}
	\caption{BER performance versus transmit power ($Q=4$).}
	\vspace{-2ex}
	\label{fig6}
\end{figure}

Fig. \ref{fig6} compares the error performance of FD-NC and FD coherent schemes, where the coherence block length is $Q=4$.
	From \eqref{eq_B}, $B=8$ bits ($B=14$ bits) are transmitted by each user per coherence block, when each FD-NC user employs $B_v=1$ bit ($B_v=2$ bits) to characterize the $v^{th}$ dimension of the cube-split constellation with $v=1,\cdots,6$.
On the other hand, the FD coherent users keep the first two symbol intervals for the pilot-based channel estimation as shown in Fig. \ref{fig2_TransmissionSchemes}. 
	Thus, the FD coherence users utilize $16$-QAM ($128$-QAM) signaling in the last two symbol intervals of coherence block to satisfy the transmission rate of $B=8$ bits ($B=14$ bits) per coherence block.
The promising numerical results show that the proposed FD-NC has lower error floor in every investigated scenarios than the FD coherent scheme. 
	Also, FD-NC achieves  $2.8$ dB and $5.8$ dB performance improvement  compared to the FD coherent for $B=8$ and $B=14$ bits, respectively.

\section{Conclusions}\label{sec_CONC}
This paper has proposed a novel full-duplex non-coherent (FD-NC) transmission scheme for mMIMO systems using analog beamforming (ABF).
	For the non-coherent communications, a structured Grassmannian constellation has been {proposed} for the symbol detection without a priori CSI.
We have developed the transmit and receive ABF via the slow time-varying AoD/AoA information. 
	Thus, the proposed ABF technique does not require the instantaneous CSI.
The numerical results imply that as the array size increases, the proposed FD-NC remarkably improves the error performance compared to both half-duplex non-coherent (HD-NC) and HD coherent schemes.
	Furthermore, compared to the FD coherent transmission, the proposed FD-NC achieves better error performance, {and} reduces the error floor observed in the high transmit power regime.

	\ifCLASSOPTIONcaptionsoff
	\newpage
	\fi
	\bibliographystyle{IEEEtran}
	\bibliography{bibAsil_2111}

\begin{thebibliography}{10}
\providecommand{\url}[1]{#1}
\csname url@samestyle\endcsname
\providecommand{\newblock}{\relax}
\providecommand{\bibinfo}[2]{#2}
\providecommand{\BIBentrySTDinterwordspacing}{\spaceskip=0pt\relax}
\providecommand{\BIBentryALTinterwordstretchfactor}{4}
\providecommand{\BIBentryALTinterwordspacing}{\spaceskip=\fontdimen2\font plus
\BIBentryALTinterwordstretchfactor\fontdimen3\font minus
  \fontdimen4\font\relax}
\providecommand{\BIBforeignlanguage}[2]{{%
\expandafter\ifx\csname l@#1\endcsname\relax
\typeout{** WARNING: IEEEtran.bst: No hyphenation pattern has been}%
\typeout{** loaded for the language `#1'. Using the pattern for}%
\typeout{** the default language instead.}%
\else
\language=\csname l@#1\endcsname
\fi
#2}}
\providecommand{\BIBdecl}{\relax}
\BIBdecl

\bibitem{Uwaechia2020}
A.~N. {Uwaechia} \emph{et~al.}, ``A comprehensive survey on millimeter wave
  communications for fifth-generation wireless networks: Feasibility and
  challenges,'' \emph{IEEE Access}, vol.~8, pp. 62\,367--62\,414, 2020.

\bibitem{Mass_MIMO_Precoding_Survey}
N.~Fatema \emph{et~al.}, ``Massive {MIMO} linear precoding: A survey,''
  \emph{IEEE Syst. J.}, vol.~12, no.~4, pp. 3920--3931, Dec. 2017.

\bibitem{Mass_MIMO_Hyb_Survey}
I.~{Ahmed} \emph{et~al.}, ``A survey on hybrid beamforming techniques in 5{G}:
  Architecture and system model perspectives,'' \emph{IEEE Commun. Surveys
  Tuts.}, vol.~20, no.~4, pp. 3060--3097, 4th Quart. 2018.

\bibitem{Analog_BF_mMIMO}
S.~Zhang \emph{et~al.}, ``{ON–OFF} analog beamforming for massive {MIMO},''
  \emph{IEEE Trans. Veh. Technol.}, vol.~67, no.~5, pp. 4113--4123, 2018.

\bibitem{FD_MIMO_ANALOG}
R.~{López-Valcarce} \emph{et~al.}, ``Analog beamforming for full-duplex
  millimeter wave communication,'' in \emph{2019 16th Int. Symp. Wireless
  Commun. Syst. (ISWCS)}, Aug. 2019, pp. 687--691.

\bibitem{FD_Survey_2019}
K.~E. {Kolodziej} \emph{et~al.}, ``In-band full-duplex technology: Techniques
  and systems survey,'' \emph{IEEE Trans. Microw. Theory Techn.}, vol.~67,
  no.~7, pp. 3025--3041, July 2019.

\bibitem{FD_Survey}
Z.~Zhang \emph{et~al.}, ``Full-duplex wireless communications: Challenges,
  solutions, and future research directions,'' \emph{Proc. IEEE}, vol. 104,
  no.~7, pp. 1369--1409, July 2016.

\bibitem{FD_AntennaSep_74dB}
E.~{Everett} \emph{et~al.}, ``Passive self-interference suppression for
  full-duplex infrastructure nodes,'' \emph{IEEE Trans. Wireless Commun.},
  vol.~13, no.~2, pp. 680--694, 2014.

\bibitem{ASIL_FD_SIC}
A.~Koc \emph{et~al.}, ``Full-duplex {mmWave} massive {MIMO} systems: A joint
  hybrid precoding/combining and self-interference cancellation design,''
  \emph{IEEE Open J. Commun. Soc.}, vol.~2, pp. 754--774, 2021.

\bibitem{nonCohMasMIMO}
H.~Xie \emph{et~al.}, ``Non-coherent massive {MIMO} systems: A constellation
  design approach,'' \emph{IEEE Trans. Wireless Commun.}, vol.~19, no.~6, pp.
  3812--3825, 2020.

\bibitem{cubeSplitConstellation}
K.-H. Ngo \emph{et~al.}, ``Cube-split: A structured grassmannian constellation
  for non-coherent {SIMO} communications,'' \emph{IEEE Trans. Wireless
  Commun.}, vol.~19, no.~3, pp. 1948--1964, 2020.

\bibitem{ASIL_ABHP_Access}
A.~{Koc} \emph{et~al.}, ``{3D} angular-based hybrid precoding and user grouping
  for uniform rectangular arrays in massive {MU-MIMO} systems,'' \emph{IEEE
  Access}, vol.~8, pp. 84\,689--84\,712, May 2020.

\bibitem{Satyanarayana2019}
K.~{Satyanarayana} \emph{et~al.}, ``Hybrid beamforming design for full-duplex
  millimeter wave communication,'' \emph{IEEE Trans. Veh. Technol.}, vol.~68,
  no.~2, pp. 1394--1404, Feb. 2019.

\bibitem{Report_5G_UMi_UMa_Rel16}
{3GPP TR 38.901}, ``5{G}: Study on channel model for frequencies from 0.5 to
  100 {GHz},'' Tech. Rep. Ver. 16.1.0, Nov. 2020.

\bibitem{ASIL_Xiaoyi_DL_CE}
X.~Zhu \emph{et~al.}, ``A deep learning and geospatial data based channel
  estimation technique for hybrid massive {MIMO} systems,'' \emph{IEEE Access},
  vol.~9, pp. 145\,115--145\,132, 2021.

\bibitem{ASIL_FD_SM_I_CSI_ICT}
A.~Koc \emph{et~al.}, ``Full-duplex spatial modulation systems under imperfect
  channel state information,'' in \emph{24th Int. Conf. Telecommun. (ICT
  2017)}, Limassol, Cyprus, May 2017.

\end{thebibliography}

\end{document}